\documentstyle[fleqn,espcrc2,epsf]{article}

\newcommand{\alt}{\mathrel{\mathop
 {\hbox{\lower0.5ex\hbox{$\sim$}\kern-0.8em\lower-0.7ex\hbox{$<$}}}}}
\newcommand\agt{\mathrel{\mathop
 {\hbox{\lower0.5ex\hbox{$\sim$}\kern-0.8em\lower-0.7ex\hbox{$>$}}}}}

\title{{\hsize=160mm
\vbox to -7pt{\vskip0cm minus100cm\normalsize\tt
Contribution to The Proceedings of the 5th International Workshop on\\
Topics in Astroparticle and Underground Physics (TAUP97)\\
Laboratori Nazionale del Gran Sasso, Italy, 7-11 September, 1997\\
edited by A.~Bottino, A.~Di Credico and P.~Monacelli
(Elsevier, Amsterdam, 1998)
\vskip1.2cm}}
Neutrino Masses in Astrophysics and Cosmology}

\author{Georg G.~Raffelt\thanks{Supported, in part, by the
Deutsche Forschungsgemeinschaft under grant No.~SFB~375.}\\
{\ }\\
Max-Planck-Institut f\"ur Physik (Werner-Heisenberg-Institut)\\
F\"ohringer Ring 6, 80805 M\"unchen, Germany}

\begin{document}

\begin{abstract}
Cosmology yields the most restrictive limits on neutrino masses and
conversely, massive neutrinos would contribute to the cosmic
dark-matter density and would play an important role for the formation
of structure in the universe. Neutrino oscillations may well solve the
solar neutrino problem and can have a significant impact on supernova
physics. The neutrino signal from a future galactic supernova could
provide evidence for cosmologically interesting neutrino masses or set
interesting limits.
\end{abstract}

\maketitle


\section{INTRODUCTION}

Within the standard model of elementary particle physics, neutrinos
play a special role in that they are the only fermions that appear
with only two degrees of freedom per family, which are massless, and
which interact only by the weak force apart from gravitation. If
neutrinos had masses or anomalous electromagnetic interactions, or if
right-handed (sterile) neutrinos existed, this would be the
long-sought ``physics beyond the standard model.'' Hence the
enthusiasm with which experimentalists search for neutrino
oscillations, neutrinoless double-beta decay, a signature for a
neutrino mass in the tritium beta decay spectrum, or for neutrino
electromagnetic dipole or transition moments.

Over the years, many speculations about hypothetical neutrino
properties and their consequences in astrophysics and cosmology have
come and gone. I shall not pursue the more exotic of those conjectures
such as strong neutrino-neu\-trino interactions by majoron and other
couplings, small neutrino electric charges, the existence of low-mass
right-handed partners to the established sequential flavors, and so
forth. Any of them can be significantly constrained by astrophysical
and cosmological methods \cite{RaffeltBook,KolbTurner}, but currently
there does not seem to be a realistic way to positively establish
physics beyond the standard model on such grounds. Therefore, I will
focus on the more conservative modifications of the standard-model
neutrino sector, namely on neutrino masses and mixings. Surely the
search for a nonvanishing mass is the quest for the holy grail of
neutrino physics!

The most important astrophysical information about neutrino properties
is the cosmological mass limit of about $40\,\rm eV$ which for
$\nu_\tau$ improves the direct experimental constraints by about six
orders of magnitude.  The only standard-model decay for mixed
neutrinos that would be fast enough to evade this limit is
$\nu_\tau\to e^+e^-\nu_e$ if $m_{\nu_\tau}>2m_e$.  However, this
channel is strongly constrained by the absence of $\gamma$-rays from
the supernova 1987A and other arguments so that a violation of the
cosmological limit requires fast invisible decays.  Additional mass
limits arise from big-bang nucleosynthesis.  Evidence for a neutrino
mass may well come from the neutrino signal of a future galactic
supernova.  Issues related to neutrino mass limits will be explored
in~Sec.~\ref{sec:masslimits}.

Currently favored models for the formation of structure in the
universe exclude neutrinos as a dark-matter candidate. Still,
with a mass of a few eV they could play an important positive
role in mixed hot plus cold dark matter scenarios and would leave an 
imprint in the power spectrum of cosmic microwave temperature 
fluctuations, topics to be discussed in Sec.~\ref{sec:darkmatter}.

The best hope for a positive identification of neutrino masses is to
discover flavor oscillations. Current indications for this phenomenon
include the solar neutrino problem, the atmospheric neutrino anomaly,
and the LSND $\bar\nu_e$ excess counts.  Because these matters have
been widely discussed in the literature I will focus in
Sec.~\ref{sec:oscillations} on neutrino oscillations in supernova
physics.  Apparently this is the only astrophysical site where flavor
oscillations could play an important direct role.


\section{MASS LIMITS}
\label{sec:masslimits}

\subsection{Cosmological Mass Limit}

Arguably the most important contribution of cosmology to neutrino
physics is the mass limit from the requirement that the universe not
be ``overclosed''~\cite{KolbTurner,Gershtein}.  In the framework of
the big-bang cosmogony one expects about as many background neutrinos
as there are microwave photons. In detail, the cosmic energy density
in massive neutrinos is found to be
$\rho_\nu=\frac{3}{11}\,n_\gamma\,\sum m_\nu$ with $n_\gamma$ the
present-day density in microwave background photons. The sum extends
over all sequential flavors. In units of the critical density
this is
\begin{equation}
\Omega_\nu h^2=\sum \frac{m_\nu}{93\,\rm eV},
\end{equation}
where $h$ is the Hubble expansion parameter in units of $100\,\rm
km\,s^{-1}\,Mpc^{-1}$. The observed age of the universe yields $\Omega
h^2\alt 0.4$ so that 
\begin{equation}
m_\nu\alt 40\,{\rm eV}
\end{equation}
for any of the known flavors.

\subsection{Decaying Neutrinos}

\begin{figure}[ht]
\centering\leavevmode
\epsfxsize=\hsize
\epsfbox{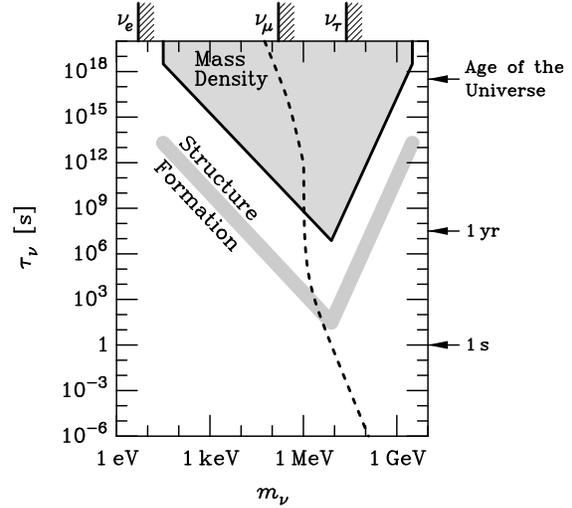}
\caption{Cosmological bounds on neutrino masses and lifetimes. The
experimental mass limits are shown above the main panel. The dashed
line is the lifetime for $\nu_\tau\to\nu_e\gamma$ and
$\nu_\tau\to\nu_e e^+ e^-$ under the assumption of maximum 
$\nu_e$-$\nu_\tau$ mixing.
\label{fig:masslifetime}}
\end{figure}

The cosmological mass limit assumes that neutrinos are stable which
most likely they are not if they have masses. Early decays into light
daughter particles would allow the energy stored in the massive
neutrinos to be redshifted enough so that the universe would not be
overclosed after all. In Fig.~\ref{fig:masslifetime} the range of
neutrino masses and lifetimes that remains forbidden is shown by the
shaded area marked ``Mass Density.'' A detailed construction of this
plot is found in Ref.~\cite{DicusKolbTeplitz}.

A decaying-neutrino cosmology actually has attractive features for
cosmic structure formation.  Standard cold dark matter produces too
much power in the density-fluctuation spectrum on small scales
(Sec.~\ref{sec:StructureFormation}).  With decaying neutrinos the
universe would become matter dominated when the massive neutrino
becomes nonrelativistic, would return to radiation domination when it
decays, and would become matter dominated again at a later time. As
structure grows by gravitational instability only in phases of matter
domination, one has two more parameters at hand (the neutrino mass and
lifetime) to tune the final density fluctuation spectrum. In the
shaded band marked ``Structure Formation'' in
Fig.~\ref{fig:masslifetime} this mechanism could help to solve the
problems of the cold dark matter cosmology~\cite{DecayingNeutrinos}.

The snag with this scenario is that within the particle-physics
standard model even massive mixed neutrinos cannot decay fast enough
because the absence of flavor-violating neutral currents prevents
processes of the sort $\nu_\tau\to\nu_e\overline\nu_e\nu_e$.  What
remains are radiative decays like $\nu_\tau\to\nu_e\gamma$ which are
of higher order and thus too slow unless one postulates interactions
beyond the standard model.  Moreover, the final-state photons would
appear as contributions to the cosmic photon backgrounds, excluding a
large range of neutrino masses and radiative lifetimes independently
of theoretical predictions~\cite{KolbTurner}.  Therefore,
decaying-neutrino cosmologies as well as a circumvention of the
cosmological mass bound require ``fast invisible decays,'' i.e.\ fast
decays with final-state neutrinos or with new particles such as
majorons. Turning this around, a violation of the cosmological mass
bound of $40\,\rm eV$ would imply physics ``far~beyond'' the standard
model.

There is one apparent exception if $m_{\nu_\tau}>2m_e$ so that
$\nu_\tau\to\nu_e e^+e^-$ is kinematically possible. Assuming maximum
$\nu_e$-$\nu_\tau$ mixing the lifetime of $\nu_\tau$ is plotted in
Fig.~\ref{fig:masslifetime} as a dashed line.  However, there are
numerous laboratory limits on the \hbox{$\nu_e$-$\nu_\tau$} mixing
angle.  Moreover, it is thought that in a supernova (SN) collapse the
gravitational binding energy of about $3{\times}10^{53}\,\rm ergs$ is
emitted almost entirely in neutrinos of all flavors.  The positrons
from the subsequent $\nu_\tau\to\nu_e e^+e^-$ decay would be trapped
in the galactic magnetic field for about $10^5\,\rm yr$ so that the
positron flux from all galactic supernovae, integrated over this time,
yields further restrictions on the decay rate~\cite{DarI}.  Finally,
the absence of a $\gamma$-ray burst in conjunction with the neutrino
signal from SN~1987A yields very restrictive limits on radiative
neutrino decays and in particular on the inner bremsstrahlung process
$\nu_\tau\to\nu_e e^+ e^-\gamma$~\cite{DarII}. Altogether, even heavy
$\tau$ neutrinos cannot escape the cosmological mass limit if masses
and mixings are the only extensions of the standard model.

\subsection{Big-Bang Nucleosynthesis}

Another mass limit arises from big-bang nucleosynthesis (BBN). The
agreement between the predicted and observed primordial light-element
abundances shows that the expansion rate in the early universe must
have been roughly the standard value, leaving little room for new
contributions. This argument excludes $1\,{\rm MeV}\alt
m_{\nu}\alt30\,{\rm MeV}$ if neutrinos live longer than about
$10^3\,\rm s$, the time of BBN~\cite{MeVstable}. If some extension of
the standard model allows them to decay on this time scale or faster
one can still derive limits in the parameter space of the mass and the
couplings which allow for the decay~\cite{MeVdecay,MeVadjust}. If the
\hbox{daughter} particles involve electron neutrinos they will modify
the $\beta$ reactions between protons and neutrons and thus change the
cosmic baryon fraction which is compatible with the observed
light-element abundances~\cite{MeVadjust}. All of this requires new
physics beyond neutrino masses and mixings and thus shall not be
pursued here any further.

\subsection{Supernova Mass Limits}

Two mass limits deserve mention which were derived from the SN~1987A
neutrino signal. First, the absence of a discernible time-of-flight
dispersion of the observed $\bar\nu_e$ burst gave rise to
$m_{\nu_e}\alt20\,\rm eV$ \cite{SNmasslimit}. There remains some
interest in this result because the laboratory limits from the tritium
$\beta$ decay endpoint spectrum seem to be plagued with unidentified
systematic errors~\cite{ParticleData}.

Second, if neutrinos had Dirac masses, helicity-flipping collisions in
the dense inner core of a SN would produce right-handed (sterile)
states which are not trapped and thus carry away the energy directly
rather than by diffusing to the neutrino sphere.  This new energy-loss
channel leads to a shortening of the expected SN~1987A $\bar\nu_e$
burst. The observed burst duration thus leads to a bound
$m_\nu({\rm Dirac}) \alt 30\,\rm keV$~\cite{RaffeltBook}.

Such a large mass violates the cosmological limit and is thus excluded
anyway unless there are nonstandard decays.  Even ``invisible''
channels typically involve left-handed final-state neutrinos which are
visible to the detectors which registered the SN~1987A signal.
Because the sterile neutrinos which escape directly from the SN core
are more energetic than those emitted from the neutrino sphere, these
events would stick out from the observed SN~1987A signal, leading to
additional limits on some decay channels~\cite{Dodelson}.

A future galactic SN would lead to much better mass limits.  A
detector like the proposed OMNIS~\cite{omnis}, which has evolved from
the former SNBO concept~\cite{snbo}, could measure neutrinos of all
flavors by a coherently enhanced neutral-current nuclear dissociation
reaction of the type \hbox{$\nu+(Z,N)\to (Z,N-1)+n+\nu$}. One could
measure time-of-flight signal dispersion
effects corresponding to neutrino masses of a
few $10\,\rm eV$ for $\nu_\mu$ or $\nu_\tau$, especially in
conjunction with the charged-current $\bar\nu_ep\to ne^+$ signal
expected for Superkamiokande~\cite{Cline}. This detector
alone would be sensitive to \hbox{$m_{\nu_e}\alt
1\,\rm eV$} on the basis of the rapid rise time of the expected
neutrino burst. Alas, galactic supernovae are rare---one expects at
most a few per century. It would still be of utmost importance to run
a supernova burst observatory for however long it takes to capture a
galactic SN neutrino signal!


\section{NEUTRINOS AS DARK MATTER}
\label{sec:darkmatter}

\subsection{Galactic Phase Space}

Massive neutrinos would seem to be natural candidates for the dark
matter which dominates the dynamics of the universe~\cite{Cowsik}.
However, they fare poorly in this regard for two main reasons.  The
first is a well-known problem with the phase space available to
neutrinos in the dark-matter haloes of galaxies (``Tremaine-Gunn
limit''~\cite{TremaineGunn}).  Neutrinos bound to the galaxy naturally
must move slower than the escape velocity $v_{\rm esc}$ so that their
momentum is bounded by $p_{\rm max}=m_\nu v_{\rm esc}$ and their
density by $n_{\rm max}=p_{\rm max}^3/3\pi^2$ due to the Pauli
exclusion principle.  The maximum local mass density in dark-matter
neutrinos is then $m_\nu n_{\rm max}= m_\nu^4 v_{\rm esc}^3/3\pi^2$,
leading to a {\it lower\/} limit on the required mass for galactic
dark-matter neutrinos of a few $10\,\rm eV$ for normal spiral
galaxies, and a few $100\,\rm eV$ for dwarf galaxies.  Therefore,
neutrinos cannot be the dark matter on all scales where it is known to
exist.

\subsection{Structure Formation}
\label{sec:StructureFormation}

\begin{figure}[ht]
\centering\leavevmode
\epsfxsize=\hsize
\epsfbox{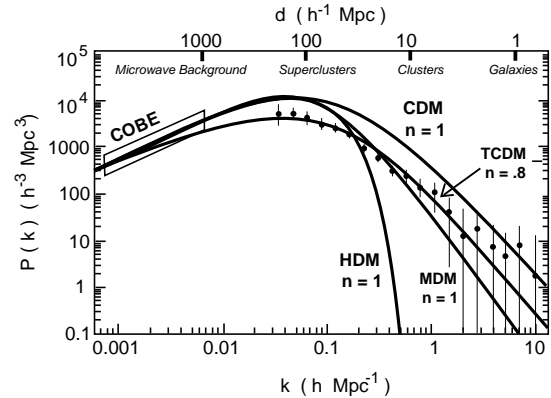}
\caption{Comparison of matter-density power spectra for cold dark
matter (CDM), tilted cold dark matter (TCDM), hot dark matter (HDM),
and mixed hot plus cold dark matter (MDM) for large-scale structure
formation \protect\cite{Steinhardt}. All theoretical curves are
normalized to COBE and include only linear approximation; nonlinear
corrections become important below about $10\,\rm Mpc$.
\label{fig:powerspectrum}}
\end{figure}

The main argument against neutrino dark matter arises from our current
understanding of how the observed structure forms in the cosmic matter
distribution.  One pictures a primordial distribution of low-amplitude
density fluctuations which are later amplified by the action of
gravity. The final distribution of galaxies depends on both the nature
of the dark matter and the original fluctuation spectrum.  This
reasoning leads to pictures like Fig.~\ref{fig:powerspectrum} where
the power spectrum of the matter distribution is shown as a function
of wave-number or length scale.  The data are derived from the
observed galaxy distribution.

Rather general arguments as well as inflationary models of the early
universe predict a roughly scale-invariant primordial fluctuation
spectrum (Harrison-Zeldovich-spectrum). One normalizes its amplitude
to the COBE observations of the spectrum of cosmic microwave
background temperature fluctuations.  Fig.~\ref{fig:powerspectrum}
reveals that with this normalization a standard cold dark matter
scenario predicts more power in the small-scale galaxy distribution
than is observed.

Neutrinos, on the other hand, represent ``hot dark matter'' (HDM)
because they stay relativistic until very late. This implies that
their collisionless streaming erases the primordial fluctuation
spectrum on small scales, suppressing the formation of small-scale
structure (Fig.~\ref{fig:powerspectrum}).  One way out is that the
original seeds for structure formation are not provided by initial
density fluctuations but rather by something like cosmic strings or
textures which cannot be erased by free
streaming~\cite{Brandenberger}. Such scenarios may or may not be
excluded at present~\cite{defects}, but they surely have been deserted
by all cosmologists apart from a few dedicated cosmic-string {\it
aficionados}.

The problem of a standard CDM cosmology depicted in
Fig.~\ref{fig:powerspectrum} can be patched up in a variety of
ways. One may tinker with the primordial fluctuation spectrum which
may have been almost, but not quite, of the Harrison-Zeldovich
form. One example of such a ``tilted cold dark matter'' (TCDM) result
is shown in Fig.~\ref{fig:powerspectrum}.

Another patch-up is to invoke a mixed hot plus cold dark matter (MDM
or CHDM) cosmology (Fig.~\ref{fig:powerspectrum}) where the hot
component erases enough of the initial power on small scales to
compensate for the overproduction by pure CDM~\cite{MDM}.  In a flat
universe ($\Omega=1$) the best fit is obtained with a total mass in
neutrinos corresponding to $\sum m_\nu=5\,\rm eV$ with an
equipartition of the masses among the flavors.

\subsection{Cosmic Microwave Background}

\begin{figure}[ht]
\centering\leavevmode
\epsfxsize=6.0cm
\epsfbox{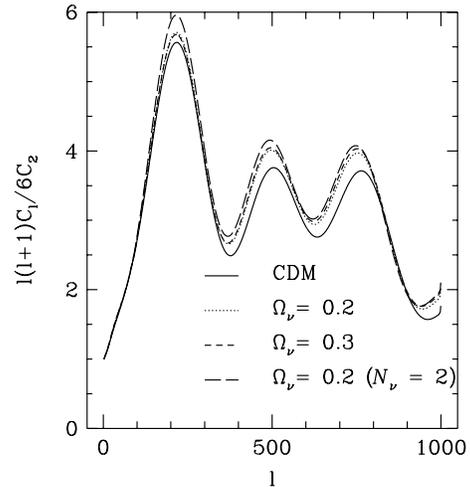}
\caption{Power spectrum of the temperature sky map for the cosmic
microwave background in a cold dark matter cosmology, and three
variants of mixed dark matter~\protect\cite{Stebbins}. 
\label{fig:cmb}}
\end{figure}

Granted that something like a CDM cosmology describes our universe,
how will we ever know if it contains a small component of neutrino
dark matter? One crucial source of information will be the precision
sky maps of the cosmic microwave background
from the future MAP and Planck Surveyor
(formely COBRAS/SAMBA) satellites~\cite{skymaps}. Such sky maps are
usually interpreted in terms of the multipole expansion of their power
spectrum. For a pure CDM cosmology the expected power as a function of
the multipole order $l$ is shown in Fig.~\ref{fig:cmb} as a solid
line. The modified power spectra for three versions of a mixed hot
plus cold dark matter cosmology are also shown.

The first ambition of current cosmic microwave experiments is to
identify the first of the ``Doppler peaks'' in the power spectrum.
However, with the high angular resolution planned for the satellite
experiments one will be able, in principle, to distinguish between the
CDM and the MDM curves of Fig.~\ref{fig:cmb}. Of course, there are
other unknown cosmological parameters such as the overall mass
density, the Hubble constant, the cosmological constant, the baryon
fraction, and so forth, which all affect the expected power spectrum.
All of them have to be determined by fitting the power spectrum to the
observations.  Therefore, it remains to be seen if a small neutrino
component of the overall dark matter density can be identified in
future cosmic microwave data.

In summary, neutrinos as a universal dark matter particle are strongly
disfavored, but with a mass of a few eV they could play a very useful
role for structure formation and as a dark matter component which is
less clustered than CDM. One may be able to identify the imprint of
this component in future cosmic microwave sky maps.


\section{NEUTRINO OSCILLATIONS}
\label{sec:oscillations}

\subsection{Current Evidence}

While neutrino masses would play a very important role in cosmology,
it appears unlikely that cosmological arguments or observations alone
will be able to prove or disprove this hypothesis anytime soon.
Therefore, the most realistic and systematic path is to search for
neutrino oscillations.  Unsurprisingly, a vast amount of experimental
effort is dedicated to this end. While there is yet no uncontestable
positive signature for oscillations, there exist a number of
experimental ``anomalies'' that are best explained by this phenomenon.

The most recent example is an experiment at Los Alamos where neutrinos
are produced in a proton beam dump. The secondary positive pions decay
according to $\pi^+\to\mu^++\nu_\mu$ and the muons according to
$\mu^+\to e^++\bar\nu_\mu+\nu_e$. In the Liquid Scintillator Neutrino
Detector (LSND) about 30 meters downstream, a significant number of
excess $\bar\nu_e$ counts was obtained which cannot be due to the
primary source but which can be interpreted as the appearance of
oscillated $\bar\nu_\mu$'s~\cite{LSND}.  If this interpretation were
correct, the $\nu_e$-$\nu_\mu$ mass difference could be of order
$1\,\rm eV$ or more, pointing to cosmologically significant neutrino
masses. At the present time one has to wait and see if more LSND data
and other experiments, notably KARMEN~\cite{karmen}, will confirm this
claim.

Another indication for oscillations arises from atmospheric
neutrinos. Their production is very similar to the LSND experiment,
except that the higher-energy cosmic-ray protons produce both
positively and negatively charged pions and kaons in roughly equal
proportions so that one expects about equally many neutrinos as
antineutrinos, and a $\nu_\mu$:$\,\nu_e$ flavor ratio of about
2:1. Some measurements agree with this prediction~\cite{Frejus}, but
several detectors have observed a flavor ratio more like 1:1
(``atmospheric neutrino anomaly'')~\cite{atmospheric}. Further,
Kamiokande has seen an angular dependence of the flavor ratio as
expected for oscillations due to the different path lengths through
the Earth from the atmosphere to the detector~\cite{Fukuda} and most
recently Superkamiokande had made a similar case~\cite{SuperK}.  These
observations can be explained by $\nu_\mu$-$\nu_e$ or
$\nu_\mu$-$\nu_\tau$ oscillations, but the former possibility is now
ruled out by the CHOOZ reactor experiment~\cite{CHOOZ}. The
$\nu_\mu$-$\nu_\tau$ oscillation interpretation requires nearly
maximum mixing with a mass difference of about $0.1\,\rm eV$.

The longest-standing hint for oscillations arises from solar
neutrinos.  The masses and mixing angles which are required to explain
the measured flux deficits in terms of oscillations have been updated
in Ref.~\cite{Hata}, but they remain close to the textbook
values~\cite{RaffeltBook}. The two MSW solutions indicate a mass
difference around $0.003\,\rm eV$ while the vacuum solution would
require about $10^{-5}\,\rm eV$.

It is well known that these indications for oscillations require three
different mass differences which are not compatible with each
other. Therefore, not all of the oscillation interpretations can be
correct unless one appeals to neutrino degrees of freedom beyond the
sequential flavors, i.e.\ to the existence of sterile
neutrinos~\cite{Caldwell}.

\subsection{Early Universe}

Meanwhile it remains of interest to look for astrophysical effects
where neutrino oscillations could be important. Neutrinos dominate the
dynamics of the early universe and so it is natural to wonder if
oscillations could be important there. However, because all flavors
are in thermal equilibrium with each other the usual flavor
oscillations would not change anything.  Oscillations into sterile
neutrinos can have nontrivial and interesting effects~\cite{BBNosci},
but following the philosophy of this presentation of mostly ignoring
everything other than masses and mixings for the sequential flavors
I will not discuss these matters here.

\subsection{Supernova Physics}

Concentrating on flavor oscillations between sequential neutrinos,
supernovae are natural sites to look for nontrivial consequences. A
type~II SN occurs when a massive star ($M\agt 8\,M_\odot$) has reached
the end of its life. It consists of a degenerate iron core, surrounded
by several shells of different nuclear burning phases. Iron cannot
gain energy by nuclear fusion so that no further burning phase can be
ignited. As the iron core grows in mass it eventually reaches its
Chandrasekhar limit of about $1.4\,M_\odot$, i.e.\ the maximum mass
that can be supported by electron degeneracy pressure. The subsequent
collapse is halted only at nuclear densities where the equation of
state stiffens, causing a shock wave to form at the edge of the inner
core. It advances outward and eventually expels the mantle and
envelope, an event which is observed as the SN explosion.  The
implosion of the core is transformed into an explosion of the outer
parts of the star by this ``shock and bounce'' mechanism.

Most of the binding energy of the newly formed compact star is
radiated away by neutrinos. The collapsed core is so hot and dense
that even neutrinos are trapped. The cooling takes several seconds
which corresponds to a neutrino diffusion time scale from the center
to the ``neutrino sphere'' where these particles can escape. It is
thought that the energy is roughly equipartitioned between all
(anti)neutrino flavors, but the spectra are different.  Various
studies find for the average expected neutrino energy~\cite{Janka93}
\begin{equation}\label{eq:energies}
\langle E_\nu\rangle=\cases{10{-}12\,{\rm MeV}&for $\nu_e$,\cr
14{-}17\,{\rm MeV}&for $\bar\nu_e$,\cr
24{-}27\,{\rm MeV}&for $\nu_{\mu,\tau}$ and 
               $\bar\nu_{\mu,\tau}$.\cr}
\end{equation}  
The differences arise from the main trapping reactions, namely $\nu_e
n \to p e^-$, $\bar\nu_e p\to n e^+$, and $\nu N\to N\nu$ with $N=n$
or $p$. The charged-current reactions have larger cross sections than
the neutral-current ones and there are more neutrons than protons so
that the $\nu_e$'s have the hardest time to escape. They emerge from
the largest radii and thus from the coldest layers.

In detail the spectra formation depends subtly on the neutrino
transport near the neutrino sphere~\cite{Janka95}. A recent scrutiny
of the neutrino interaction rates suggests that the spectral energies
may be less different between the flavors than had been
thought~\cite{spectra}, but a self-consistent implementation is not
yet available.

It is conceivable that (resonant) oscillations occur outside of the
neutrino sphere so that the spectra between two flavors are
swapped. This would affect the explosion mechanism itself which does
not work quite as simple as described above. Because the shock wave
forms within the core it has to move through a layer of iron before
reaching the stellar mantle. By dissociating iron it loses energy and
stalls after a few $100\,\rm ms$ in typical calculations. The
deposition of energy by neutrinos which emerge from the inner core is
thought to revive the shock so that it resumes its outward
motion. However, this ``delayed explosion mechanism'' still does not
seem to work in typical calculations because the transfer of energy
to the shock wave is not efficient enough.

\begin{figure}[t]
\centering\leavevmode
\epsfxsize=7cm
\epsfbox{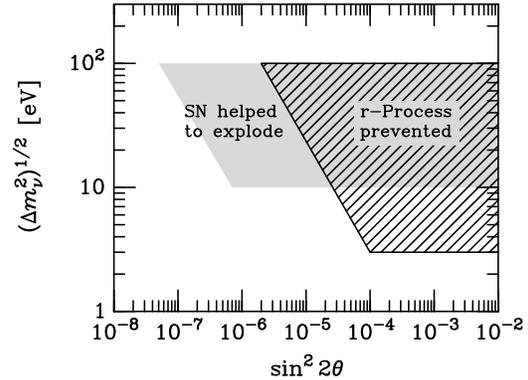}
\caption{Mixing parameters between $\nu_e$ and $\nu_\mu$ or $\nu_\tau$
where a spectral swap by resonant oscillations would be efficient
enough to help explode supernovae (schematically after
Ref.~\protect\cite{Fuller92}), and where it would prevent r-process
nucleosynthesis (schematically after Ref.~\protect\cite{Qian}).
\label{fig:snoscillations}}
\end{figure}

If neutrinos follow a ``normal'' mass hierarchy with $\nu_e$ dominated
by the lightest mass eigenstate one can have MSW oscillations between,
say, $\nu_e$ and $\nu_\tau$. If this occurs between the neutrino
sphere and the stalling shock wave the $\nu_e$'s arriving there
really are oscillated $\nu_\tau$'s and thus have the higher spectral
energies characteristic for that flavor, leading to a more effective
energy transfer to the shock wave \cite{Fuller92}. Because the MSW
transition must occur close to the neutrino sphere where the matter
densities are large, neutrino mass differences in the $10\,\rm eV$
regime are required (Fig.~\ref{fig:snoscillations}).

Oscillations may also affect the r-process synthesis of heavy elements
(neutron capture) which may well occur in the high-entropy ``hot
bubble'' in a SN between the neutron star and the advancing shock wave
a few seconds after collapse.  Because $\langle
E_{\nu_e}\rangle<\langle E_{\bar\nu_e}\rangle$ the $\beta$ processes
shift the neutrino-driven wind to the required neutron-rich phase.
However, if oscillations cause a spectral swap between, say, $\nu_e$
and $\nu_\tau$ this energy hierarchy is inverted and the wind is
shifted to a proton-rich phase, preventing the occurrence of the
r-process~\cite{Qian}. Because this argument applies to a later phase
than the explosion argument above, the neutron star has thermally
settled so that the matter gradients at its surface are much
steeper. This makes it harder to meet the adiabaticity condition,
reducing the range of mixing angles where the MSW effect operates
(Fig.~\ref{fig:snoscillations}).

At the present time it is not certain if r-process nucleosynthesis
indeed occurs in supernovae so that the hatched are in
Fig.~\ref{fig:snoscillations} cannot be taken as an exclusion plot.
More importantly, there is a range of mixing parameters below the
hatched region where the spectral swap is only partial and causes an
{\it increase\/} of the neutron fraction, actually helping the
r-process~\cite{fuller97}.
 
\subsection{Pulsar Recoils}

Neutron stars (pulsars) usually have strong magnetic fields. They
cause the neutrino refractive index to be anisotropic so that the MSW
resonance would not occur at precisely the same radius everywhere in
the SN core. As a result the total neutrino luminosity may not be
precisely isotropic, causing a small recoil of the newborn neutron
star~\cite{Kusenko}. It is not clear, however, if one can actually
achieve the 1--2\% anisotropy which is required to explain the
observed pulsar velocities of around $500\,\rm
km\,s^{-1}$~\cite{Lorimer}.

\subsection{SN~1987A Signal Interpretation}

Oscillations would modify the SN~1987A neutrino signal, notably the
``prompt $\nu_e$ burst'' which precedes the main cooling phase. It
arises when the shock wave breaks through the surface of the iron
core, suddenly releasing $\nu_e$'s by the reactions $e^-p\to n\nu_e$
from a layer encompassing perhaps a few $0.1\,M_\odot$.  In the IMB
and Kamiokande water Cherenkov detectors which registered the SN~1987A
signal the $\nu_e$-$e$ scattering reaction could have produced
forward-peaked electrons as a signature for this burst, in agreement
with the first event in Kamiokande, although one would have expected
only a fraction of an event.

\begin{figure}[b]
\centering\leavevmode
\epsfxsize=6cm
\epsfbox{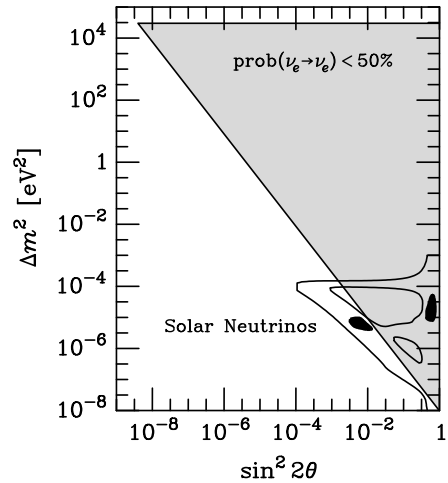}
\caption{Mixing parameters between $\nu_e$ and $\nu_\mu$ or $\nu_\tau$
where the prompt $\nu_e$ burst would have resonantly oscillated into
another flavor (after Ref.~\protect\cite{Dirk}). A normal mass
hierarchy is assumed where $\nu_e$ is dominated by the lightest mass
eigenstate.  For orientation, the Kamiokande solar MSW triangle and
the MSW solutions to the solar neutrino problem are also shown.
\label{fig:dirk}}
\end{figure}

Resonant oscillations would have transformed the $\nu_e$ burst into
$\nu_\mu$'s or $\nu_\tau$'s which have a much smaller scattering cross
section on electrons. In Fig.~\ref{fig:dirk} the shaded triangle shows
the mixing parameters for which the oscillation probability in the
stellar mantle and envelope would have exceeded 50\%.  The small-angle
MSW solution is not in conflict with the interpretation that the first
SN~1987A Kamiokande event was indeed from $\nu_e$-$e$ scattering.  Of
course, this single event does not lead to the opposite conclusion
that the large-angle MSW solution was ruled~out.

Most of the 19 events must have been due to the $\bar\nu_e p\to n e^+$
reaction.  For a normal mass hierarchy resonant oscillations cannot
swap the $\bar\nu_e$ spectra with another flavor; they can affect
only the $\nu_e$ spectrum.  Therefore, the observed events represent
the original  $\bar\nu_e$ source spectrum unless the mixing angle
is large enough to allow for significant nonresonant oscillations.
Large mixing angles are motivated by the large-angle MSW and the
vacuum solution to the solar neutrino problem as well as the
oscillation interpretation of the atmospheric neutrino anomaly.

\begin{figure}[b]
\centering\leavevmode
\vbox{
\epsfxsize=6cm
\epsfbox{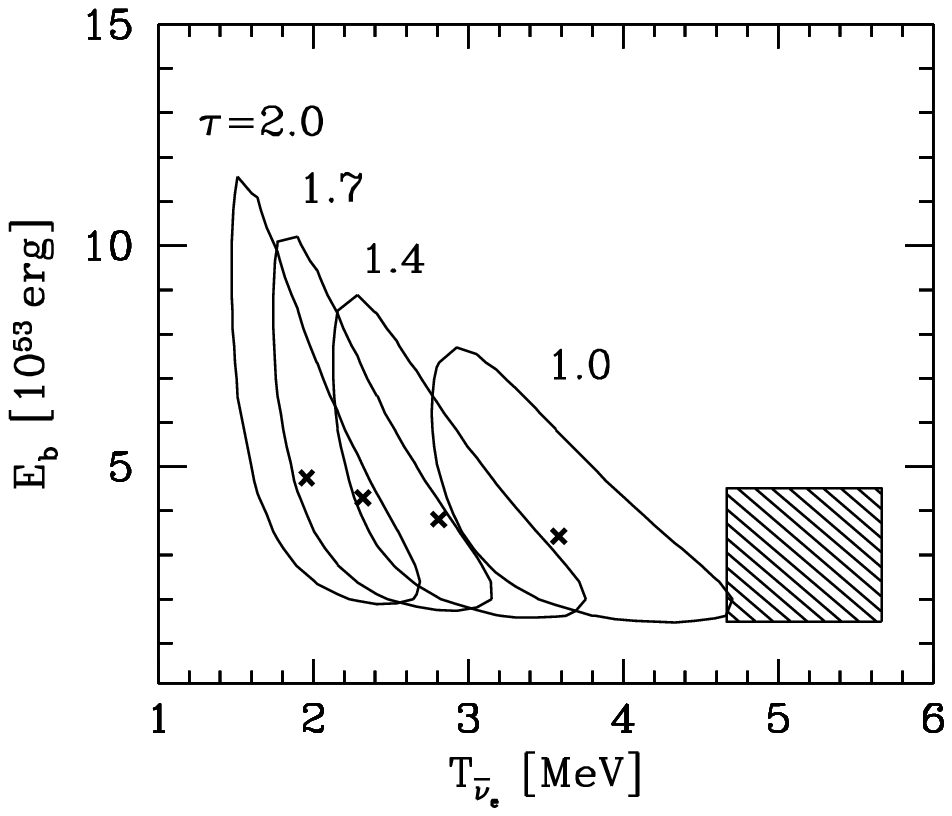}
\vskip0cm
\epsfxsize=6cm
\epsfbox{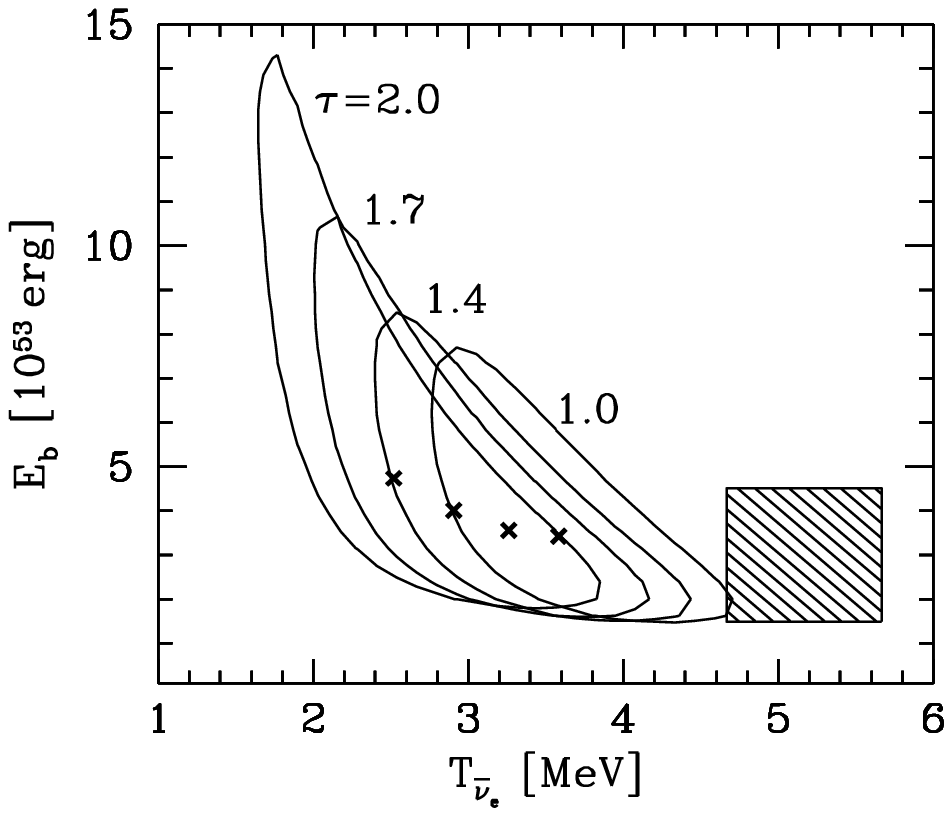}
}
\caption{Confidence contours (95\%) for the neutron star binding
energy and temperature of the primary $\bar\nu_e$ spectrum for the
given values of $\tau=T_{\bar\nu_\mu}/T_{\bar\nu_e}$
\protect\cite{Jegerlehner}. The hatched area is the range of
theoretical predictions.
{\it Upper Panel}:
Neutrino mixing parameters of
the solar vacuum solution ($\Delta m^2=10^{-10}\,\rm eV^2$,
$\sin^22\Theta=1$).
{\it Lower Panel}:
Large-angle MSW solution
($\Delta m^2=10^{-5}\,\rm eV^2$, $\sin^22\Theta=0.8$).
\label{fig:swap}}
\end{figure}

One way of interpreting the observed SN~1987A events is to derive
best-fit values for the total binding energy $E_{\rm b}$ and the
spectral temperature of the observed $\bar\nu_e$'s which is defined by
$T_{\bar\nu_e}=\frac{1}{3}\langle E_{\bar\nu_e}\rangle$. For certain
mixing parameters and relative spectral temperatures $\tau\equiv
T_{\bar\nu_\mu}/T_{\bar\nu_e}$ the results from such an
analysis~\cite{Jegerlehner} are shown in Fig.~\ref{fig:swap}.  For
$\tau=1$ oscillations have no effect; this is identical to the
no-oscillation scenario.

According to Eq.~(\ref{eq:energies}) a typical value for the relative
spectral temperature is $\tau=1.7$. This is inconsistent with the
vacuum solution to the solar neutrino problem
(Fig.~\ref{fig:swap}). The expected event energies in the detector
would have been even larger than in the standard case, contrary to the
relatively low energies that were actually observed. Put another way,
if the vacuum solution to the solar neutrino problem is borne out by
future experiments, there is a serious conflict between the SN~1987A
observations and current theoretical predictions.

For the large-angle MSW solution the conflict is less severe
(Fig.~\ref{fig:swap}). In this case the flavor evolution is adiabatic
in the SN envelope so that propagation eigenstates emerge from the
surface. On the path through the Earth to the detectors matter-induced
``regeneration effects'' partly restore the original source spectra,
reducing the overall impact of neutrino oscillations.

In summary, the SN~1987A neutrino observations disfavor the
large-angle solutions to the solar neutrino problem, even though the
data are too sparse to reach this conclusion ``beyond reasonable
doubt.''


\section{DISCUSSION AND SUMMARY}
\label{sec:conclusions}

In the absence of any compelling theoretical reason for neutrinos to
be strictly massless it is commonplace to assume that they do carry
small masses and that the flavors mix. Cosmology provides by far the
most restrictive limit of about $40\,\rm eV$ on the mass of all
sequential flavors. This limit cannot be circumvented by decays unless
neutrinos interact by new forces which allow for ``fast invisible''
(i.e.\ nonradiative) channels. Therefore, a neutrino mass in excess of
the cosmological limit would signify that either the particle-physics
or the cosmological standard model require nontrivial revisions.

Neutrinos are unfavored dark matter candidates because of the
well-known problems of a hot dark matter cosmology.  The cold
dark matter picture works impressively well even though it appears to
overproduce structure on small scales. This problem can be patched up
by a number of different modifications, one of them being a hot plus
cold dark matter scenario with a neutrino component corresponding to
$m_{\nu_e}+m_{\nu_\mu}+m_{\nu_\tau}\approx 5\,\rm eV$. However, it
looks unlikely that this sort of scenario can be unambiguously
identified by cosmological methods alone. Even the most ambitious
future cosmic microwave sky maps will have a hard time to identify
this model unambiguously in view of the remaining uncertainty in other
cosmological parameters.

Depending on the exact mixing parameters, neutrino oscillations have
severe consequences for supernova physics and the signal
interpretation of SN~1987A or a future galactic supernova. Especially
for neutrino masses in the cosmologically interesting regime,
oscillations can affect the explosion mechanism and r-process
nucleosynthesis. However, the current understanding of SN physics is
too uncertain and the SN~1987A data are too sparse to tell if neutrino
oscillations are either required or excluded. Still, it remains
fascinating that a neutrino mass as small as a few eV has {\it any\/}
significant consequences outside of cosmology.

Even though massive neutrinos may play an important role in cosmology
and supernova physics, realistically we will know if this is indeed
the case only by more direct measurements.  The most promising
approach is by oscillation experiments.  Already, oscillations can
explain the atmospheric neutrino anomaly, the LSND $\bar\nu_e$ excess
counts, and the solar neutrino problem even though a simultaneous
explanation of all three phenomena is not possible by flavor
oscillations between sequential neutrinos alone.

If neutrinos do have masses and if their mass differences are as small
as suggested by solar and atmospheric neutrinos, then a cosmological
role is only possible if all three mass eigenvalues are in the eV
range and almost equal.  The common scale of these quasi-degenerate
masses cannot be determined in oscillation experiments. Therefore, it
remains of utmost importance to push tritium $\beta$ decay and
neutrinoless $\beta\beta$ decay experiments below the $1\,\rm eV$
threshold for $m_{\nu_e}$.  Moreover, it remains of utmost importance
to measure the neutrino light curve from a future galactic supernova
with a high-statistics experiment such as Superkamiokande or the
proposed OMNIS~\cite{omnis}.  Besides a wealth of other information
one would be sensitive to an eV mass for $\nu_e$ and to the $10\,\rm
eV$ scale for $m_{\nu_\mu}$ and $m_{\nu_\tau}$.  Galactic supernovae
may be rare, but the scientific harvest would be worth the wait!


\end{document}